\documentclass[ALICE,manyauthors]{cernphprep}

\usepackage[comma,square,numbers,sort&compress]{natbib}
\usepackage{hyperref}
\usepackage{lineno}
\usepackage{graphicx}
\usepackage{booktabs}
\usepackage[T1]{fontenc}

\newcommand{\pbar}{$\rm{\overline{p}}$\xspace}

\newcommand{\pt}{\ensuremath{p_{\rm T}}\xspace}

\newcommand{\pbpb}{Pb--Pb\xspace}
\newcommand{\sNN}{$\sqrt{s_{\mathrm{NN}}} = $\xspace}
\newcommand{\hyptriton}{$^{3}_{\Lambda}\mathrm{H}$\xspace}
\newcommand{\antihyptriton}{$^{3}_{\overline{\Lambda}}\mathrm{\overline{H}}$\xspace}
\newcommand{\elio}{$^{3}\mathrm{He}$\xspace}
\newcommand{\antielio}{$^{3}\mathrm{\overline{He}}$\xspace}

\begin{document}%

\begin{titlepage}
\PHyear{2019}
\PHnumber{148}      
\PHdate{10 July}  
%

\title{\hyptriton and \antihyptriton lifetime measurement in \pbpb collisions at \newline \sNN 5.02 TeV via two-body decay}
\ShortTitle{\hyptriton and \antihyptriton lifetime in \pbpb collisions at \sNN 5.02 TeV }   

\Collaboration{ALICE Collaboration\thanks{See Appendix~\ref{app:collab} for the list of collaboration members}}
\ShortAuthor{ALICE Collaboration} 

\begin{abstract}
An improved value for the  lifetime of the (anti-)hypertriton has been obtained using   the data sample of \pbpb collisions at \sNN 5.02 TeV collected by the ALICE experiment at the LHC. 
The \newline(anti-)hypertriton has been reconstructed via its charged two-body mesonic decay channel and the lifetime has been determined from an exponential fit to the d$N$/d($ct$) spectrum. 
The measured value, $\tau$ = 242$^{+34}_{-38}$ (stat.) $\pm$ 17 (syst.) ps, is compatible with representative theoretical predictions, thus contributing to the solution of the longstanding hypertriton lifetime puzzle.
\end{abstract}
\end{titlepage}
\setcounter{page}{2}

\section{Introduction}
Hypernuclei are bound states of nucleons and hyperons and they are mainly produced by means of ($K^{-},\pi^{-}$), ($\pi^{+},K^{+}$) and ($e$,$e^{'}K^{+}$) reactions on stable nuclear targets \cite{Botta2012, Gal:2016boi}. 
Hypernuclei are particularly interesting because they can be used as experimental probes for the study of the hyperon-nucleon (Y--N) interaction. The knowledge of this interaction has become more relevant in recent years due to its connection to the modelling of astrophysical objects like neutron stars \cite{Lattimer:2004pg,SchaffnerBielich:2008kb}. In the inner core of  neutron stars, the creation of hyperons is energetically favoured compared to a purely nucleonic matter composition \cite{tolos_centelles_ramos_2017}. The presence of hyperons as  additional degrees of freedom leads to a considerable softening of the matter equation of state (EOS). The resulting EOS inhibits the formation of large mass neutron stars. This is incompatible with the observation of neutron stars as heavy as two solar masses \cite{Lattimer:2004pg},  constituting what is referred to as the ``hyperon puzzle''. Many attempts were made to solve this puzzle, e.g. by introducing three-body forces leading to an additional repulsion that can counterbalance the large gravitational pressure and  allow for larger star masses. To constrain the parameter space of such models, a detailed knowledge of the Y--N interaction  and of the three-body Y--N--N interaction is mandatory, including $\Lambda$, $\Sigma$ and $\Xi$ states.
The lifetime of a hypernucleus depends on the strength of the Y--N interaction, and therefore a precise determination of the lifetime of hypernuclei provides information on the Y--N interaction strength \cite{Dalitz:1962eb, Kamada:1997rv}.

The recent observation of hypernuclei and the determination of their lifetimes in experiments with relativistic heavy ion collisions has triggered a particular interest. All the results published so far are related to the lightest hypernucleus, the hypertriton \hyptriton, which is a bound state formed by a proton, a neutron and a $\Lambda$, and its charge conjugate the anti-hypertriton \antihyptriton.
The results have been obtained at the Relativistic Heavy Ion Collider (STAR experiment) \cite{Abelev:2010rv}, at the SIS18 (HypHI Collaboration) \cite{Rappold:2013fic} and at the Large Hadron Collider (ALICE Collaboration) \cite{adam:2016h3l}. 

The separation energy of the $\Lambda$ in this hypernucleus is only about 130 keV \cite{Davis:2005mb}, which results in an RMS radius (average distance of the $\Lambda$ to the deuteron) of 10.6 fm \cite{Braun-Munzinger:2018hat, Hildenbrand:2019sgp}. A very low binding energy implies a small change of the wave function of the $\Lambda$ in a nucleus and hence one can expect the lifetime of the hypertriton to be very close to that of the free $\Lambda$ hyperon ($\tau_\Lambda = (263.2 \pm 2.0)$ ps~\cite{PhysRevD.98.030001}).

Early hypertriton lifetime measurements were done with imaging techniques (i.e. emulsions, bubble chambers) and the results are lower than or consistent with the value of the free $\Lambda$ lifetime \cite{PhysRev.136.B1803, PhysRevLett.20.819, PhysRev.180.1307, Bohm:1970se, Keyes:1970ck, Keyes:1974ev}. However, most of the measurements performed with these techniques are based on very small samples of events, thus resulting in a large statistical uncertainty. The recent measurements of the lifetime of \mbox{(anti-)\hyptriton} produced in ultra-relativistic heavy-ion collisions or in relativistic ion fragmentation \cite{Chen:2018tnh}, even though affected by statistical and systematic uncertainties bigger than 10$\%$, are in agreement among each other and are lower than the free $\Lambda$ lifetime \cite{adam:2016h3l,Rappold:2013fic,Adamczyk:2017buv}. 

However, the few existing theoretical calculations predict that the lifetime of the \hyptriton should be very close to the lifetime of free $\Lambda$. The most comprehensive \hyptriton lifetime calculation is from Rayet and Dalitz \cite{Dalitz:1966}; they obtained an estimate in the range from 239.3--255.5 ps. More recent calculations from Congleton \cite{Congleton:1992kk} and Kamada et al. \cite{Kamada:1997rv} yield a value of 232 ps and 256 ps, respectively.   
This scenario stimulated, in the last years, a new interest from both experimentalists and theoreticians for more precise measurements of the \hyptriton lifetime.

In this letter, the lifetime of the (anti-)\hyptriton measured in \mbox{\pbpb} collisions at \sNN 5.02 TeV by the ALICE experiment is presented.
In Section~\ref{sec_alice}, the ALICE detector is briefly described. The details of the data sample, analysis technique and systematic uncertainties are presented in Section~\ref{sec_std},where also a new analysis approach to crosscheck the results is introduced in the subsection~\ref{sec_unbfit}. Finally the result is compared with previous measurements and with theoretical predictions in Section~\ref{sec_conclusion}. 

\section{The ALICE apparatus} \label{sec_alice}
A detailed description of the ALICE apparatus and data acquisition framework can be found in \cite{Aamodt:2008zz, Abelev:2014ffa}.
The main detectors used in this analysis are the V0 detector, the Inner Tracking System (ITS) and the Time Projection Chamber (TPC), which are located inside a solenoid creating a magnetic field of 0.5 T. 
The V0 detector \cite{Abbas:2013taa} consists of two arrays of scintillator counters (V0A and V0C), placed around the beam-pipe on both sides of the interaction region. They cover the pseudorapidity ranges 2.8 $< \eta <$ 5.1 and $-3.7 < \eta < -1.7$, respectively. The V0 detector is used to define the Minimum Bias (MB) trigger, which is characterized by a coincidence signal in the V0A and in the V0C, and to determine the centrality of the collisions \cite{PhysRevLett.116.222302}.
The ITS \cite{Aamodt:2010aa} is the closest detector to the interaction point within ALICE. It is composed of six layers of silicon detectors, with radii between 3.9 and 43 cm from the interaction point. The six layers use three different technologies: silicon pixel detector (SPD), silicon drift detector (SDD) and silicon strip detector (SSD). The ITS has full azimuthal coverage $0 \leq \varphi \leq 2\pi$ and covers the pseudorapidity range $|\eta| < $ 0.9.
The TPC \cite{ALME2010316} is a gaseous detector, mainly used for tracking and for particle identification (PID) via the specific energy loss (d$E$/d$x$), with a total sensitive volume of 90 m$^3$ filled with a mixture of 88\% Ar and 12\% CO$_2$. 
The reconstructed clusters in TPC and ITS are the starting point of the track finder algorithm, which adopts the Kalman filter technique \cite{Kalman::1987}. These tracks are used to determine the primary collision vertex with a precision better than 50 $\mu$m in the plane transverse to the colliding beams \cite{Abelev:2014ffa}.

\section{Data sample and analysis technique}
\label{sec_std}
In this letter, the lifetime of the (anti-)hypertriton is determined by exploiting the 2-body mesonic decay channel with charged pions, namely \hyptriton $\rightarrow$ \elio + $\pi^-$ and \antihyptriton $\rightarrow$ \antielio + $\pi^+$. Both \hyptriton and \antihyptriton candidates are used for this measurement.

The analysis is performed using the data sample of \pbpb collisions at \sNN 5.02 TeV collected by the ALICE experiment at the end of 2015. To ensure a uniform acceptance and reconstruction efficiency in the pseudorapidity region $|\eta|<$ 0.9, only those events are selected whose reconstructed primary vertex was within $\pm$10 cm from the nominal position of the interaction point along the beam axis. The analysed sample contains approximately 90 million events in the centrality interval 0-90$\%$.

The \hyptriton and \antihyptriton identification is based on the topology of their weak decays and on the reconstruction of the tracks of their decay products, referred to as daughter particles. 
The weakly decaying hypernuclei are reconstructed using the algorithm which was previously used for the K$^0_{\rm S}$ and $\Lambda$ production analyses \cite{Abelev:2013xaa} and which is typically adopted for a 2-body weak decay topology.
At first, the algorithm uses the TPC and ITS clusters to reconstruct the daughter tracks and then combines them in order to obtain a V-shaped decay vertex. More details on this algorithm can be found in \cite{Aamodt:2011zza,Abelev:2014ffa}.

The daughter tracks are selected in the pseudorapidity region $|\eta|<$ 0.9 and are required to have at least 70 clusters out of 159 in the TPC, in order to guarantee a resolution $\sigma$ better than 5$\%$ on track momentum and of about 6$\%$ for the d$E$/d$x$ \cite{Abelev:2014ffa}. Moreover, the $\chi^2$ per TPC cluster is required to be less than 5 and tracks with kink topologies are rejected. The particle identification (PID) of the daughters (\elio, \antielio, $\pi^{\pm}$) is performed following the method described in \cite{Aamodt:2011zza}, which is used in many analyses of the ALICE Collaboration. It is based on the difference between the measured and the expected d$E$/d$x$ for a selected particle species normalized to the energy loss resolution in the detector, $\sigma$ for short, and is referred to as the n$\sigma$ method in this letter.
In particular, an $|$n$\sigma| \leq $ 3 is required, in a track-by-track approach, with respect to the expected $\pi$ and \elio specific energy loss in the TPC. The pions can be identified up to a momentum of about 1.2 GeV/\textit{c}, beyond which there is considerable contamination from kaons and protons. The \elio, having a charge of $z = 2e$, can be identified cleanly up to 7 GeV/\textit{c}. The \elio is also produced in the detector material due to spallation. These are produced at low transverse momenta, as reported by the ALICE experiment \cite{Adam:2015vda}. As a consequence the \elio candidate is required to have a transverse momentum (\pt) greater than 1.8 GeV/\textit{c}, where the spallation processes are negligible.  

The \hyptriton and \antihyptriton candidates are selected by applying topological and kinematic selection criteria on the decay products. The distance of closest approach (DCA) between the two daughter tracks and the DCA of $\pi^{\pm}$ tracks from the primary vertex are required to be lower than 0.7 cm and larger than 0.1 cm respectively. The candidates are selected whose cosine of the angle between the total momentum of the daughter tracks at the secondary vertex and the vector connecting the primary and secondary vertex (pointing angle) is larger than 0.995. Two additional selections on the \hyptriton and \antihyptriton rapidity ($|y| <$ 0.8) and transverse momentum (2 $<$ \pt $<$ 9 GeV/\textit{c}) are applied. All the selection criteria previously described have been studied with a dedicated Monte Carlo production, in order to improve the background rejection, and are summarized in Table~\ref{tab_cuts_yields_ctspectra}.

\begin{table}[!h]
	\centering
	\caption{Selection criteria applied for the identification of the daughter candidate tracks and for the reconstruction of the hypertriton candidate.}
	\begin{tabular}{lc}
	\toprule
	\multicolumn{2}{c}{\textbf{Selection criteria}}\\
	\hline
	\multicolumn{2}{l}{Track selections}\\
	\hline
	$|\eta|$ & $<$ 0.9 \\
	Number of TPC clusters & $>$ 70 \\
	$\chi^2$ per TPC cluster & $<$ 5 \\
	Kink topology & Rejected \\
	$|$n$\sigma|$ for TPC PID & $\leq$ 3 \\
	\hline
	\multicolumn{2}{l}{Daughter candidate selections}\\
	\hline
	$\pi$ \pt (GeV/\textit{c}) & 0.2-1.2 \\
	DCA between $\pi^{\pm}$ and primary vertex (cm) & $>$ 0.1 \\
	\elio \pt (GeV/\textit{c}) & $\geq$ 1.8 \\
	DCA$_{\mathrm{tracks}}$ (cm) & $<$ 0.7 \\
	\hline
	\multicolumn{2}{l}{Hypertriton candidate selections} \\
	\hline
	cos($\theta_{\mathrm{pointing}}$) & $\geq$ 0.995 \\
	$|$\textit{y}$|$ & $\leq$ 0.8\\
	\pt (GeV/\textit{c}) & 2--9 \\
	\bottomrule
	\end{tabular}
	\label{tab_cuts_yields_ctspectra}
\end{table}

\begin{figure}[!htb]
	\centering
	\includegraphics[scale=0.73]{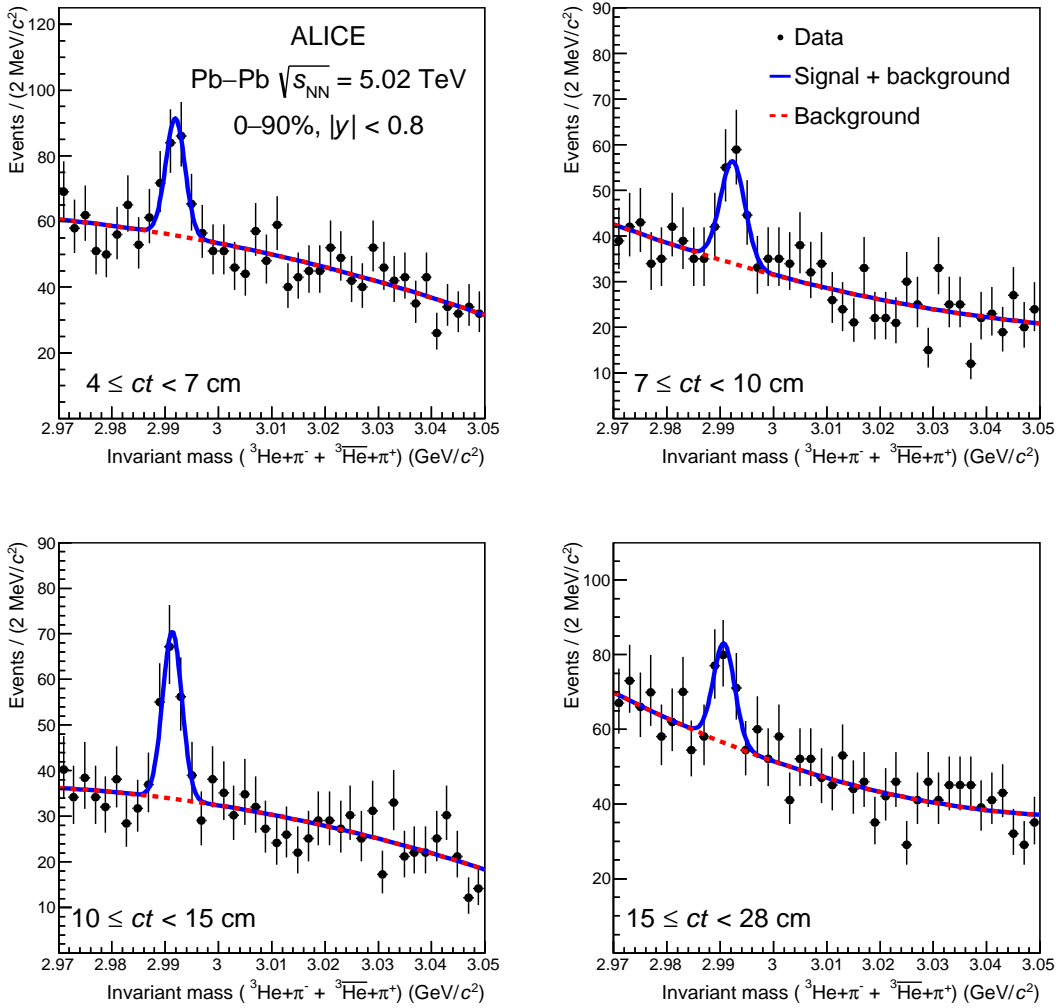}
	\caption{Invariant mass distribution of (\elio, $\pi^-$) and (\antielio, $\pi^+$) for the four $ct$ intervals used to determine the \hyptriton and \antihyptriton lifetime. The solid blue curve represents the function used to perform the fit and the red dashed curve represents the background component.}
	\label{fig1}
\end{figure}

The sample of \hyptriton and \antihyptriton candidates is divided in four $ct$ = $MLc$/$p$ intervals for the lifetime determination, where $c$ is the speed of light, $t$ is the proper time of the candidate, $M$ is the mass of the candidate, $L$ is the decay distance and $p$ is the reconstructed momentum. The mass $M$ of the hypertriton is obtained from the measured values of mass of p, n and $\Lambda$ \cite{PhysRevD.98.030001} and of the binding energy \cite{Davis:2005mb}, and has been fixed at $M$ = 2.99116 $\pm$ 0.00005 GeV/$c^2$. 
The four $ct$ intervals are 4 $\leq c$t $<$ 7 cm, 7 $\leq c$t $<$ 10 cm, 10 $\leq c$t $<$ 15 cm and 15 $\leq c$t $<$ 28 cm. The corresponding invariant mass distributions are shown in Fig.~\ref{fig1} and are fitted, in each $ct$ interval, with a function which is the sum of a Gaussian, used to interpolate the signal, and a second order polynomial, used to describe the background. The fit is performed using the maximum-likelihood estimate and the fit function is represented as a solid blue line.

From the fit, the mean values $\mu$ and the widths $\sigma$ of each distribution are extracted. In particular, the signal width is in the range \mbox{1.7--2.1 MeV/$c^2$}, depending on the $ct$ interval, and is driven by the detector resolution. The raw yield of the signal is defined as the integral of the Gaussian function in a $\pm$ 3$\sigma$ region around the mean value above the background. The significance of the signal in the four $ct$ intervals varies in the range 3.1--4.9.

The yield is corrected in each $ct$ bin for the detector acceptance, the reconstruction efficiency and the absorption of the \hyptriton (\antihyptriton) in the detector material.
The efficiency$\times$acceptance is determined with a dedicated Monte Carlo simulation, where the \hyptriton and \antihyptriton are injected on top of a HIJING event  \cite{hijing:1991} and are allowed to decay into charged two-body and three-body final states. The simulated particles are propagated through the ALICE detectors using the GEANT3 transport code \cite{brun:1082634} and then reconstructed following the same procedure as adopted for the data.

\begin{figure}[!htb]
	\centering
	\includegraphics[scale=0.55]{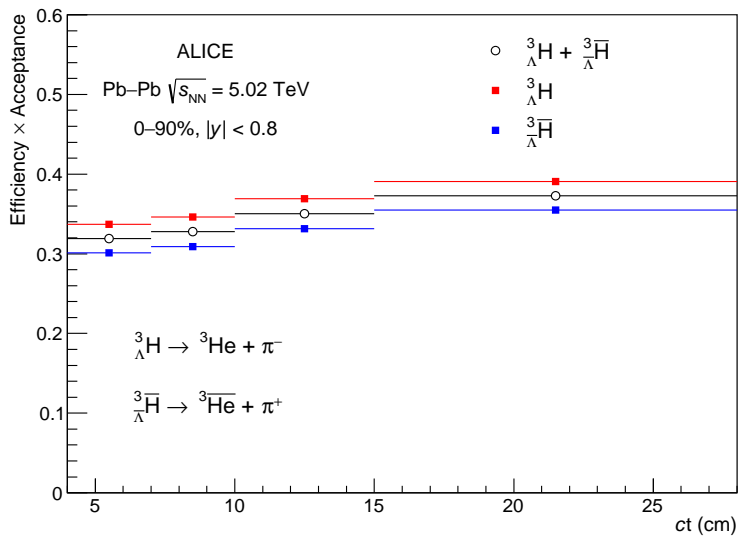}
	\caption{Efficiency$\times$acceptance as a function of $ct$ for \hyptriton (red square), \antihyptriton (blue square) and \hyptriton+\antihyptriton (black open circle) in the same $ct$ intervals selected for the raw yields extraction.}
	\label{fig2}
\end{figure}

The aforementioned transport code does not properly describe the interactions of the (anti-)(hyper-)nuclei with the material of the apparatus. Thus, a correction factor for the absorption of \hyptriton (\antihyptriton) and \elio (\antielio) is estimated, based on the p (\pbar) absorption probability measured in the ALICE detector \cite{Abbas2013}. The usage of this experimental measurement offers the advantage of taking automatically into account the cross section and the effective material of the detector crossed by a charged particle. The same absorption probability for protons and neutrons has been assumed and the \elio(\antielio) has been considered as a state of three independent p (\pbar) as verified in \cite{adam:2016h3l}. The absorption probability, computed as the third power of that of one proton, goes from 11$\%$ at low \pt to 6$\%$ at high \pt for \antielio while it is constant at 6$\%$ for \elio.
The evaluation of the \hyptriton (\antihyptriton) absorption probability is done following the same approach. However, to take into account the small $\Lambda$ separation energy (B$_{\Lambda}$ = 0.13 $\pm$ 0.05 MeV \cite{Davis:2005mb}), the \hyptriton absorption cross-section is increased by 50$\%$ with respect to the one of the \elio \cite{EVLANOV1998624,KOX198515}, as described in the ALICE measurement in \pbpb collisions at \sNN 2.76 TeV \cite{adam:2016h3l}. This leads to an absorption probability between 16$\%$ and 9$\%$ for \antihyptriton as a function of \pt while it is constant at 9$\%$ for \hyptriton. The correction factor to be applied is:
\begin{equation}
	k = k_{\mathrm{abs},^{3}_{\Lambda}\mathrm{H}} + (1-k_{\mathrm{abs},^{3}_{\Lambda}\mathrm{H}})k_{\mathrm{abs},^{3}\mathrm{He}}
\end{equation}
where $k_{\mathrm{abs},^{3}_{\Lambda}\mathrm{H}}$ is the probability that the \hyptriton is absorbed between the primary and the secondary vertex while $k_{\mathrm{abs},^{3}\mathrm{He}}$ is the probability that the daughter \elio is absorbed between the secondary vertex and the TPC inner wall.
For each $ct$ interval, the efficiency x acceptance has been calculated using the absorption corrected numbers of reconstructed \hyptriton and \antihyptriton. Fig.~\ref{fig2} shows the efficiency$\times$acceptance (black marker) which is used for the lifetime determination and is obtained by combining \hyptriton and \antihyptriton after the absorption correction is applied. This distribution is also shown separately for \hyptriton and \antihyptriton and the difference is due to the absorption correction which is bigger for the anti-matter. 

The main sources of systematic uncertainties on each $ct$ bin used for the lifetime evaluation are the absorption correction, the single track efficiency and the uncertainty on the detector material budget.
The systematic uncertainty on the absorption correction is mainly due to the assumption used for the \hyptriton (\antihyptriton) cross-section. This uncertainty is evaluated by varying this assumption between a lower and an upper limit. The first one is obtained by setting the \hyptriton (\antihyptriton) cross-section equal to the \elio(\antielio) absorption cross-section and the second one as twice the \elio(\antielio) absorption cross-section. This leads to an uncertainty of 5.2$\%$ for each $ct$ interval, as reported in Table \ref{tab_1}.

The second source of systematic uncertainty is related to the material budget description in the simulation. An uncertainty on the knowledge of the ALICE detector material budget of 4.5$\%$ was determined in a previous study \cite{Abelev:2014ffa}. The systematic uncertainty is estimated using two dedicated Monte Carlo productions, varying the material budget accordingly, and amounts to 1$\%$ for the yields in all $ct$ intervals.

The systematic uncertainty due to the single-track efficiency and the different choices of the track quality selections has been investigated \cite{Acharya:2018qsh} and amounts to 4$\%$. For the analysis of the two-body decay of \hyptriton an uncertainty of 8$\%$ is assumed in all $ct$ intervals. The summary of the systematic uncertainties is reported in Table \ref{tab_1}, where the total uncertainty is obtained as sum in quadrature of each contribution of the individual sources.

\begin{figure}[!h]
	\centering
	\includegraphics[scale=0.6]{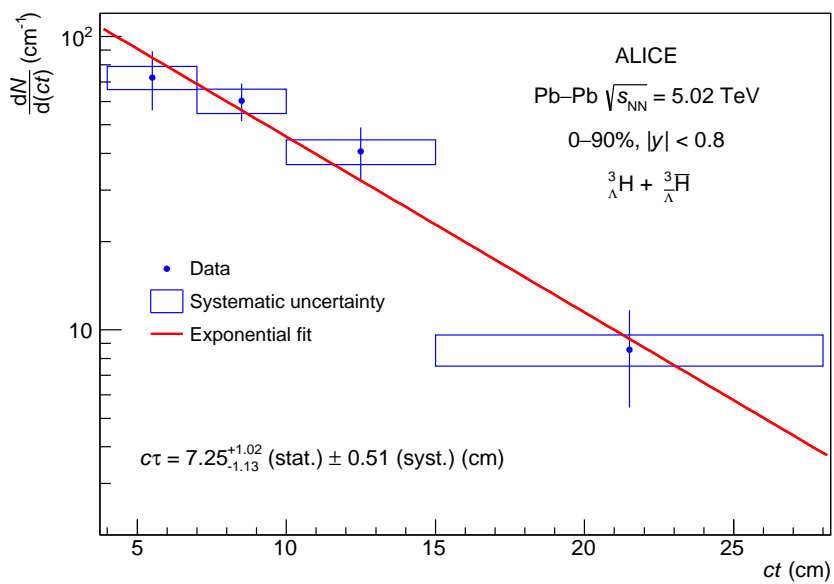}
	\caption{Corrected d$N$/d($ct$) spectrum fitted with an exponential function (red line) used to extract the (\hyptriton + \antihyptriton) lifetime. The bars and boxes represent the statistical and systematic uncertainties, respectively.}
	\label{fig3}
\end{figure}

\begin{table}[!h]
	\centering
	\caption{Summary of the systematic uncertainties used in the lifetime analysis. The total uncertainty assigned in each \textit{c}t interval is the sum in quadrature of the single sources.}
	\begin{tabular}{lr}
	\hline
	\multicolumn{2}{c}{\textbf{Systematic uncertainties}}\\
	\hline
	Source	& Value \\
	\hline	
	Absorption & 5.2$\%$ \\
	Material budget & 1$\%$ \\
	Single track efficiency & 8$\%$\\
	\hline
	\textbf{Total} & 9.5$\%$ \\
	\hline
	\end{tabular}
	\label{tab_1}
\end{table}
The corrected d$N$/d($ct$) spectrum is shown in Fig.~\ref{fig3} where the blue markers are the corrected yield with their statistical uncertainty, while the boxes represents the systematic uncertainty.\newline
The lifetime is determined with an exponential fit (red line) and the slope results in a proper decay length of \mbox{$c\tau$ = 7.25$^{+1.02}_{-1.13}$ (stat.) $\pm$ 0.51 (syst.) cm}, corresponding to a lifetime $\tau$ = 242$^{+34}_{-38}$ (stat.) $\pm$ 17 (syst.) ps. The systematic uncertainty for the lifetime value is determined by assuming the systematic uncertainties in each $ct$ interval as uncorrelated. 

\subsection{Unbinned fit method for lifetime extraction} \label{sec_unbfit}
In order to enforce the result described in Sec.~\ref{sec_std}, an additional analysis on the same data sample has been carried out that relies on a two-dimensional (invariant mass vs. $ct$) unbinned fit approach.
The method can be summarized in three steps: i) fit to the $ct$-integrated invariant mass distribution; ii) tune the function used to describe the combinatorial background; iii) fit to the $ct$ distribution with a function which is the sum of three exponentials, one to describe the signal and two to describe the background.

The first step is performed with a function that is the sum of a Gaussian, for the signal, and a second order polynomial, for the background. 
The mean value $\mu$ and the $\sigma$ are 2.9913$\pm$0.0004 GeV/$c^2$ and 0.0020$\pm$0.0005 GeV/$c^2$ respectively and are used to define the boundaries of the signal region and the sidebands, which correspond to the intervals $\mu$ $\pm$ 3$\sigma$ and $\pm$ 3$\sigma$ to $\pm$ 9$\sigma$ with respect to the mean value, respectively.

The second step consists in fitting the $ct$ distribution of the background in the sidebands using a function that is the sum of two exponentials. The fit is performed simultaneously in the two sideband regions with the R{\footnotesize OO}F{\footnotesize IT} package \cite{roofit:2008}. 
The result is then used as background parameterization for the fit in the signal region. 

\begin{figure}[!t]
	\centering
	\includegraphics[scale=0.55]{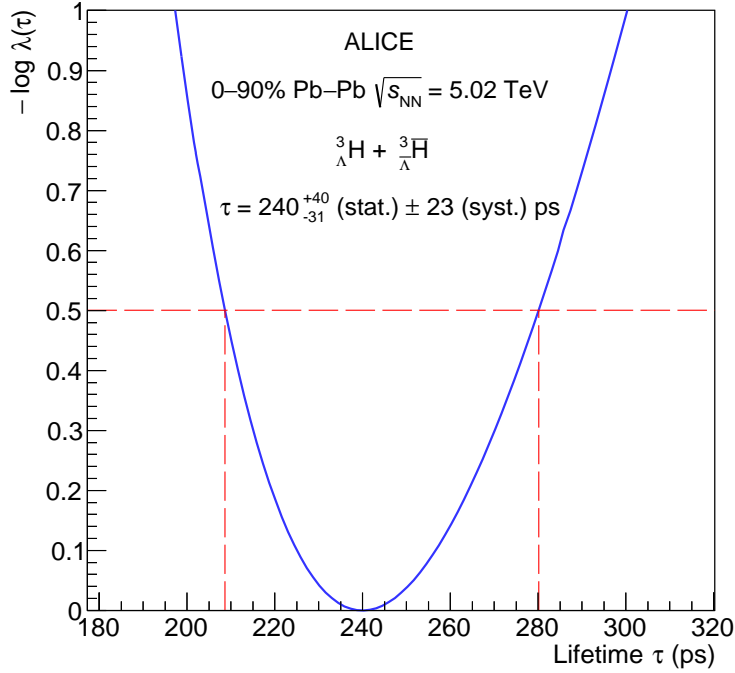}
	\caption{Lifetime value $\tau$ determined from the minimization of the log-likelihood ratio --log($\lambda(\tau)$). The statistical uncertainty is evaluated at a confidence level of 68$\%$ (red dashed lines) with the log-likelihood ratio (blue line).}
	\label{fig4}
\end{figure}

The (\hyptriton + \antihyptriton) lifetime measurement is obtained by performing the unbinned fit to the $ct$ distribution in the signal region. The total probability density function used for the fit is the sum of the two exponentials (background) and the exponential adopted to reproduce the signal. 
Since the $ct$ distribution is unbinned, the efficiency$\times$acceptance correction, evaluated as described in Sec.~\ref{sec_std}, is parametrized with a polynomial plus an exponential and it is used to scale the signal function. The observed signal distribution is described as the product of the function used for the signal and the efficiency parametrization. Thus, the lifetime is obtained with the unbinned maximum-likelihood estimate (MLE) fit to the $ct$ distribution, performed in the signal region, leading to a value of $\tau$ = 240$^{+40}_{-31}$ (stat.) $\pm$ 18 (syst.) ps, as reported in Fig.~\ref{fig4}. The statistical uncertainty of the measurement is assessed by providing the interval of the estimated $\tau$	 \cite{wilks1938}, at a confidence level of 68$\%$, which is represented by the red dashed lines, based on the log-likelihood ratio (log$\lambda(\tau)$), shown as a blue line. The result corresponds to a proper decay length \mbox{$c\tau$ = 7.20$^{+1.20}_{-0.93}$ (stat.) $\pm$ 0.54 (syst.) cm}. The sources of systematic uncertainties are the same as described in Sec.~\ref{sec_std} (Table \ref{tab_1}) and contribute to a total systematic uncertainty of 9.5$\%$ on the estimated lifetime.

The value obtained with this approach is in good agreement within 1$\sigma$ with the lifetime estimation obtained with the method described in Sec.~\ref{sec_std}, which we consider as the final value for the \hyptriton lifetime. 
Additional figures and details for the unbinned fit method are presented in \cite{alipub2019}.

\section{Discussion and conclusions}\label{sec_conclusion}
\begin{figure}[!htb]
	\centering
	\includegraphics[scale=0.87]{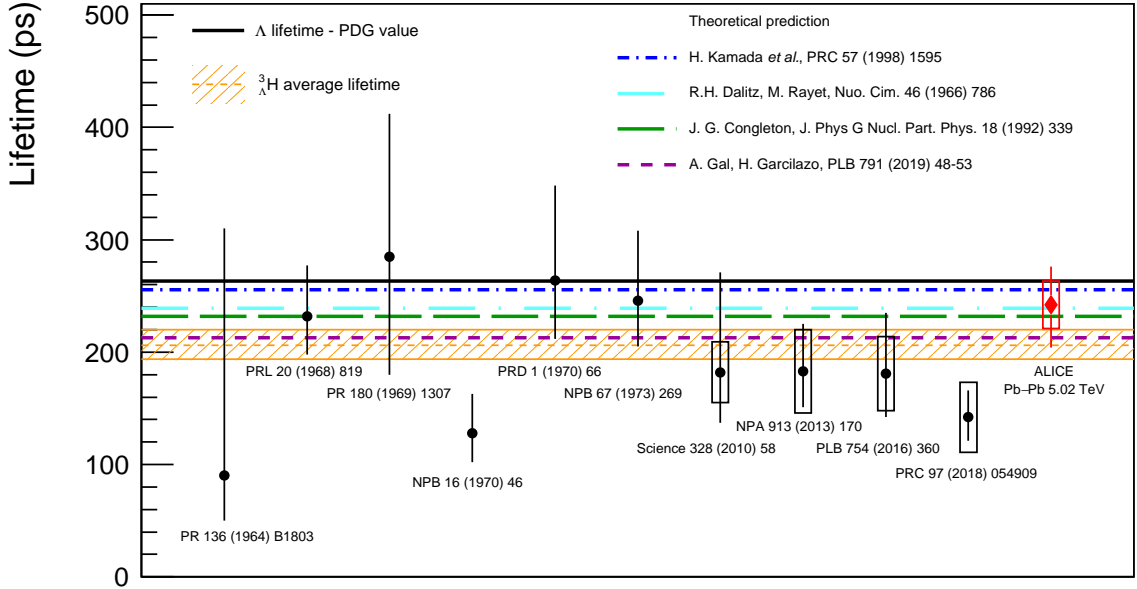}
	\caption{Collection of the \hyptriton lifetime measurements obtained with different experimental techniques. The vertical lines and boxes are the statistical and systematic uncertainties respectively. The orange band represents the average of the lifetime values and the lines at the edge correspond to 1$\sigma$ uncertainty. The dashed-dotted lines are four theoretical predictions.}
	\label{fig7}
\end{figure}

Thanks to the large data sample of heavy-ion collisions at \sNN 5.02 TeV provided by the LHC and to the excellent tracking and particle identification performance of the ALICE apparatus we have determined a precise value for the \hyptriton lifetime. The measured $\tau$ = 242$^{+34}_{-38}$ (stat.) $\pm$ 17 (syst.) ps is shown as a full red diamond in Fig.~\ref{fig7} together with other experimental results and theoretical estimates.

Early experiments \cite{PhysRev.136.B1803, PhysRevLett.20.819, PhysRev.180.1307, Bohm:1970se, Keyes:1970ck, Keyes:1974ev} were performed with visualizing techniques, namely photographic emulsion and \elio filled bubble chambers, where the tracks formed due to passage of charged particles were recorded visually. Most of the results obtained using these techniques had large uncertainties due to the limited size of the data sample at disposal. Furthermore, these measurements prevented a definite conclusion on the agreement with the theoretical predictions, which foresee a lifetime close to the value of the free $\Lambda$ hyperon. It is worthwhile to note that the small binding energy of the hypertriton makes the $\Lambda$ spend most of the time far from the deuteron core thereby not affecting the lifetime due to Y-N interaction.

The recent determination of the lifetime $\tau$ of (anti-)\hyptriton of 182$^{+89}_{-45}$ (stat.) $\pm$ 27 (syst.) ps, measured for the first time in Au--Au collisions via two-body decay by the STAR experiment at RHIC \cite{Abelev:2010rv}, revived the interest for a more precise determination of the lifetime. The HypHI Collaboration at GSI reported a value of $\tau$ = 183$^{+42}_{-32}$ (stat.) $\pm$ 37 (syst.) ps \cite{Rappold:2013fic}, which was obtained by studying the projectile fragmentation of $^{6}$Li at 2 AGeV on a carbon target.
Very recently, the ALICE experiment at the LHC measured a lifetime value \mbox{$\tau$ = 181$^{+54}_{-38}$ (stat.) $\pm$ 33 (syst.) ps \cite{adam:2016h3l}} using the data from \pbpb collisions at \mbox{\sNN 2.76 TeV} and the invariant mass analysis of the two-body decay channel. 
The average value of all results available up to 2016 was $\tau$ = 215$^{+18}_{-16}$ ps \cite{adam:2016h3l}, much lower than the theoretical estimates, motivating the need for a measurement with improved precision.
The STAR Collaboration  performed a new analysis \cite{Adamczyk:2017buv} combining the two-body and the three-body decay channels using the data sample of the RHIC beam energy scan, resulting in an even lower value of $\tau$ = 142$^{+24}_{-21}$ (stat.) $\pm$ 29 (syst.) ps.
The ALICE Collaboration exploited the data collected in \pbpb collisions at \mbox{\sNN 5.02 TeV} to carry out a new analysis of the two-body decay channel, reported in this letter. These two most recent values are reported in Fig.~\ref{fig7}. The new measurement by STAR yields a very low value as compared to the lifetime of the free $\Lambda$, while the result presented in this paper is in agreement with the theoretical predictions and it is characterized by an improved precision with respect to previous experiments. This value is also in agreement with a previous ALICE result \cite{adam:2016h3l} obtained by analyzing the data sample of \pbpb collisions at \sNN 2.76 TeV.

Besides the experimental results, the theoretical predictions for the \hyptriton lifetime are reported in Fig.~\ref{fig7} for comparison with the data. The calculation performed by Dalitz and Rayet \cite{Dalitz:1966}, represented with a dot-long dashed cyan line, took into account the phase space factors and the Pauli principle, including also corrections to account for  final state pion scattering and the non-mesonic weak decay channel. More recently, a prediction for the \hyptriton lifetime quite close to the one of the free $\Lambda$ hyperon was published by Congleton \cite{Congleton:1992kk} (dashed green line in Fig.~\ref{fig7}), obtained using updated values for N--N and Y--N potentials. The prediction by Kamada et al. \cite{Kamada:1997rv} (dotted-dashed blue line) was performed with a rigorous determination of the hypernucleus wave function and of the three nucleons scattering states, thus finding a value of 256 ps, which is the closest to the free $\Lambda$ lifetime value. Recently, Garcilazo and Gal performed a calculation \cite{Gal:2018bvq} using the wave function generated by solving three-body Faddeev equations and adding the final-state interactions of the pions. Their prediction of 213 ps is shown as a dotted purple line.   

A statistical combination of all the experimental results, including the most recent values determined by the STAR and the ALICE experiment, leads to a world average of $\tau$ = 206$^{+15}_{-13}$ ps for the \hyptriton lifetime and is represented with an orange band in Fig. \ref{fig7}. The method used for this evaluation is the same as described in \cite{adam:2016h3l}.
Furthermore world averages were calculated grouping the measurements on the basis of the experimental techniques, obtaining \mbox{$\tau_{visual}$ = 224$^{+23}_{-20}$ ps} and \mbox{$\tau_{\rm{HI}}$ = 189$^{+22}_{-20}$ ps} for the visualizing techniques and the heavy-ion experiments, respectively. These values are consistent and in agreement, also with the world average, and this suggests that the results are not affected by the technique used for the measurement.

Despite the addition of two recent high precision measurements of the \hyptriton lifetime, one well below and the other closer to the theoretical predictions, the situation has hardly changed with the current world average, now more than 3 $\sigma$ below the lifetime of the free $\Lambda$ hyperon.
In the future a very large data sample will be collected with heavy-ion collisions during LHC Run 3 (2021-2023) and Run 4 (2027-2029) \cite{Citron:2018lsq}. At the end of Run 4, ALICE expects to reduce the statistical uncertainty on the lifetime down to 1$\%$ and significantly improve the systematic uncertainty, which at present is 9.5$\%$. Furthermore, it would be beneficial in view of a more solid comparison with the theoretical predictions, to have new measurements performed at lower energies at RHIC and SIS and by using different experimental techniques at the J-PARC and MAMI facilities.
A measurement of the lifetime to a precision of a few percent will guide and constrain the theoretical input leading to a more precise determination of the Y-N interaction, eventually contributing to solving the hyperon puzzle.

\newenvironment{acknowledgement}{\relax}{\relax}
\begin{acknowledgement}
\section*{Acknowledgements}

The ALICE Collaboration would like to thank all its engineers and technicians for their invaluable contributions to the construction of the experiment and the CERN accelerator teams for the outstanding performance of the LHC complex.
The ALICE Collaboration gratefully acknowledges the resources and support provided by all Grid centres and the Worldwide LHC Computing Grid (WLCG) collaboration.
The ALICE Collaboration acknowledges the following funding agencies for their support in building and running the ALICE detector:
A. I. Alikhanyan National Science Laboratory (Yerevan Physics Institute) Foundation (ANSL), State Committee of Science and World Federation of Scientists (WFS), Armenia;
Austrian Academy of Sciences, Austrian Science Fund (FWF): [M 2467-N36] and Nationalstiftung f\"{u}r Forschung, Technologie und Entwicklung, Austria;
Ministry of Communications and High Technologies, National Nuclear Research Center, Azerbaijan;
Conselho Nacional de Desenvolvimento Cient\'{\i}fico e Tecnol\'{o}gico (CNPq), Universidade Federal do Rio Grande do Sul (UFRGS), Financiadora de Estudos e Projetos (Finep) and Funda\c{c}\~{a}o de Amparo \`{a} Pesquisa do Estado de S\~{a}o Paulo (FAPESP), Brazil;
Ministry of Science \& Technology of China (MSTC), National Natural Science Foundation of China (NSFC) and Ministry of Education of China (MOEC) , China;
Croatian Science Foundation and Ministry of Science and Education, Croatia;
Centro de Aplicaciones Tecnol\'{o}gicas y Desarrollo Nuclear (CEADEN), Cubaenerg\'{\i}a, Cuba;
Ministry of Education, Youth and Sports of the Czech Republic, Czech Republic;
The Danish Council for Independent Research | Natural Sciences, the Carlsberg Foundation and Danish National Research Foundation (DNRF), Denmark;
Helsinki Institute of Physics (HIP), Finland;
Commissariat \`{a} l'Energie Atomique (CEA), Institut National de Physique Nucl\'{e}aire et de Physique des Particules (IN2P3) and Centre National de la Recherche Scientifique (CNRS) and R\'{e}gion des  Pays de la Loire, France;
Bundesministerium f\"{u}r Bildung und Forschung (BMBF) and GSI Helmholtzzentrum f\"{u}r Schwerionenforschung GmbH, Germany;
General Secretariat for Research and Technology, Ministry of Education, Research and Religions, Greece;
National Research, Development and Innovation Office, Hungary;
Department of Atomic Energy Government of India (DAE), Department of Science and Technology, Government of India (DST), University Grants Commission, Government of India (UGC) and Council of Scientific and Industrial Research (CSIR), India;
Indonesian Institute of Science, Indonesia;
Centro Fermi - Museo Storico della Fisica e Centro Studi e Ricerche Enrico Fermi and Istituto Nazionale di Fisica Nucleare (INFN), Italy;
Institute for Innovative Science and Technology , Nagasaki Institute of Applied Science (IIST), Japan Society for the Promotion of Science (JSPS) KAKENHI and Japanese Ministry of Education, Culture, Sports, Science and Technology (MEXT), Japan;
Consejo Nacional de Ciencia (CONACYT) y Tecnolog\'{i}a, through Fondo de Cooperaci\'{o}n Internacional en Ciencia y Tecnolog\'{i}a (FONCICYT) and Direcci\'{o}n General de Asuntos del Personal Academico (DGAPA), Mexico;
Nederlandse Organisatie voor Wetenschappelijk Onderzoek (NWO), Netherlands;
The Research Council of Norway, Norway;
Commission on Science and Technology for Sustainable Development in the South (COMSATS), Pakistan;
Pontificia Universidad Cat\'{o}lica del Per\'{u}, Peru;
Ministry of Science and Higher Education and National Science Centre, Poland;
Korea Institute of Science and Technology Information and National Research Foundation of Korea (NRF), Republic of Korea;
Ministry of Education and Scientific Research, Institute of Atomic Physics and Ministry of Research and Innovation and Institute of Atomic Physics, Romania;
Joint Institute for Nuclear Research (JINR), Ministry of Education and Science of the Russian Federation, National Research Centre Kurchatov Institute, Russian Science Foundation and Russian Foundation for Basic Research, Russia;
Ministry of Education, Science, Research and Sport of the Slovak Republic, Slovakia;
National Research Foundation of South Africa, South Africa;
Swedish Research Council (VR) and Knut \& Alice Wallenberg Foundation (KAW), Sweden;
European Organization for Nuclear Research, Switzerland;
National Science and Technology Development Agency (NSDTA), Suranaree University of Technology (SUT) and Office of the Higher Education Commission under NRU project of Thailand, Thailand;
Turkish Atomic Energy Agency (TAEK), Turkey;
National Academy of  Sciences of Ukraine, Ukraine;
Science and Technology Facilities Council (STFC), United Kingdom;
National Science Foundation of the United States of America (NSF) and United States Department of Energy, Office of Nuclear Physics (DOE NP), United States of America.
\end{acknowledgement}

\bibliographystyle{utphys}   
\bibliography{biblio}

\newpage
\appendix
\section{The ALICE Collaboration}
\label{app:collab}

\begingroup
\small
\begin{flushleft}
S.~Acharya\Irefn{org141}\And 
D.~Adamov\'{a}\Irefn{org93}\And 
S.P.~Adhya\Irefn{org141}\And 
A.~Adler\Irefn{org73}\And 
J.~Adolfsson\Irefn{org79}\And 
M.M.~Aggarwal\Irefn{org98}\And 
G.~Aglieri Rinella\Irefn{org34}\And 
M.~Agnello\Irefn{org31}\And 
N.~Agrawal\Irefn{org10}\textsuperscript{,}\Irefn{org48}\textsuperscript{,}\Irefn{org53}\And 
Z.~Ahammed\Irefn{org141}
% ...
\And
\renewcommand\labelenumi{\textsuperscript{\theenumi}~}

\section*{Affiliation notes}
\renewcommand\theenumi{\roman{enumi}}
\begin{Authlist}
\item \Adef{org*}Deceased
\item \Adef{orgI}Dipartimento DET del Politecnico di Torino, Turin, Italy
\item \Adef{orgII}M.V. Lomonosov Moscow State University, D.V. Skobeltsyn Institute of Nuclear, Physics, Moscow, Russia
\item \Adef{orgIII}Department of Applied Physics, Aligarh Muslim University, Aligarh, India
\item \Adef{orgIV}Institute of Theoretical Physics, University of Wroclaw, Poland
\end{Authlist}

\section*{Collaboration Institutes}
\renewcommand\theenumi{\arabic{enumi}~}
\begin{Authlist}
\item \Idef{org1}A.I. Alikhanyan National Science Laboratory (Yerevan Physics Institute) Foundation, Yerevan, Armenia
\item \Idef{org2}Bogolyubov Institute for Theoretical Physics, National Academy of Sciences of Ukraine, Kiev, Ukraine
\item \Idef{org3}Bose Institute, Department of Physics  and Centre for Astroparticle Physics and Space Science (CAPSS), Kolkata, India
\item \Idef{org4}Budker Institute for Nuclear Physics, Novosibirsk, Russia
\item \Idef{org5}California Polytechnic State University, San Luis Obispo, California, United States
\item \Idef{org6}Central China Normal University, Wuhan, China
\item \Idef{org7}Centre de Calcul de l'IN2P3, Villeurbanne, Lyon, France
\item \Idef{org8}Centro de Aplicaciones Tecnol\'{o}gicas y Desarrollo Nuclear (CEADEN), Havana, Cuba
\item \Idef{org9}Centro de Investigaci\'{o}n y de Estudios Avanzados (CINVESTAV), Mexico City and M\'{e}rida, Mexico
\item \Idef{org10}Centro Fermi - Museo Storico della Fisica e Centro Studi e Ricerche ``Enrico Fermi', Rome, Italy
\item \Idef{org11}Chicago State University, Chicago, Illinois, United States
\item \Idef{org12}China Institute of Atomic Energy, Beijing, China
\item \Idef{org13}Chonbuk National University, Jeonju, Republic of Korea
\item \Idef{org14}Comenius University Bratislava, Faculty of Mathematics, Physics and Informatics, Bratislava, Slovakia
\item \Idef{org15}COMSATS University Islamabad, Islamabad, Pakistan
\item \Idef{org16}Creighton University, Omaha, Nebraska, United States
\item \Idef{org17}Department of Physics, Aligarh Muslim University, Aligarh, India
\item \Idef{org18}Department of Physics, Pusan National University, Pusan, Republic of Korea
\item \Idef{org19}Department of Physics, Sejong University, Seoul, Republic of Korea
\item \Idef{org20}Department of Physics, University of California, Berkeley, California, United States
\item \Idef{org21}Department of Physics, University of Oslo, Oslo, Norway
\item \Idef{org22}Department of Physics and Technology, University of Bergen, Bergen, Norway
\item \Idef{org23}Dipartimento di Fisica dell'Universit\`{a} 'La Sapienza' and Sezione INFN, Rome, Italy
\item \Idef{org24}Dipartimento di Fisica dell'Universit\`{a} and Sezione INFN, Cagliari, Italy
\item \Idef{org25}Dipartimento di Fisica dell'Universit\`{a} and Sezione INFN, Trieste, Italy
\item \Idef{org26}Dipartimento di Fisica dell'Universit\`{a} and Sezione INFN, Turin, Italy
\item \Idef{org27}Dipartimento di Fisica e Astronomia dell'Universit\`{a} and Sezione INFN, Bologna, Italy
\item \Idef{org28}Dipartimento di Fisica e Astronomia dell'Universit\`{a} and Sezione INFN, Catania, Italy
\item \Idef{org29}Dipartimento di Fisica e Astronomia dell'Universit\`{a} and Sezione INFN, Padova, Italy
\item \Idef{org30}Dipartimento di Fisica `E.R.~Caianiello' dell'Universit\`{a} and Gruppo Collegato INFN, Salerno, Italy
\item \Idef{org31}Dipartimento DISAT del Politecnico and Sezione INFN, Turin, Italy
\item \Idef{org32}Dipartimento di Scienze e Innovazione Tecnologica dell'Universit\`{a} del Piemonte Orientale and INFN Sezione di Torino, Alessandria, Italy
\item \Idef{org33}Dipartimento Interateneo di Fisica `M.~Merlin' and Sezione INFN, Bari, Italy
\item \Idef{org34}European Organization for Nuclear Research (CERN), Geneva, Switzerland
\item \Idef{org35}Faculty of Electrical Engineering, Mechanical Engineering and Naval Architecture, University of Split, Split, Croatia
\item \Idef{org36}Faculty of Engineering and Science, Western Norway University of Applied Sciences, Bergen, Norway
\item \Idef{org37}Faculty of Nuclear Sciences and Physical Engineering, Czech Technical University in Prague, Prague, Czech Republic
\item \Idef{org38}Faculty of Science, P.J.~\v{S}af\'{a}rik University, Ko\v{s}ice, Slovakia
\item \Idef{org39}Frankfurt Institute for Advanced Studies, Johann Wolfgang Goethe-Universit\"{a}t Frankfurt, Frankfurt, Germany
\item \Idef{org40}Gangneung-Wonju National University, Gangneung, Republic of Korea
\item \Idef{org41}Gauhati University, Department of Physics, Guwahati, India
\item \Idef{org42}Helmholtz-Institut f\"{u}r Strahlen- und Kernphysik, Rheinische Friedrich-Wilhelms-Universit\"{a}t Bonn, Bonn, Germany
\item \Idef{org43}Helsinki Institute of Physics (HIP), Helsinki, Finland
\item \Idef{org44}High Energy Physics Group,  Universidad Aut\'{o}noma de Puebla, Puebla, Mexico
\item \Idef{org45}Hiroshima University, Hiroshima, Japan
\item \Idef{org46}Hochschule Worms, Zentrum  f\"{u}r Technologietransfer und Telekommunikation (ZTT), Worms, Germany
\item \Idef{org47}Horia Hulubei National Institute of Physics and Nuclear Engineering, Bucharest, Romania
\item \Idef{org48}Indian Institute of Technology Bombay (IIT), Mumbai, India
\item \Idef{org49}Indian Institute of Technology Indore, Indore, India
\item \Idef{org50}Indonesian Institute of Sciences, Jakarta, Indonesia
\item \Idef{org51}INFN, Laboratori Nazionali di Frascati, Frascati, Italy
\item \Idef{org52}INFN, Sezione di Bari, Bari, Italy
\item \Idef{org53}INFN, Sezione di Bologna, Bologna, Italy
\item \Idef{org54}INFN, Sezione di Cagliari, Cagliari, Italy
\item \Idef{org55}INFN, Sezione di Catania, Catania, Italy
\item \Idef{org56}INFN, Sezione di Padova, Padova, Italy
\item \Idef{org57}INFN, Sezione di Roma, Rome, Italy
\item \Idef{org58}INFN, Sezione di Torino, Turin, Italy
\item \Idef{org59}INFN, Sezione di Trieste, Trieste, Italy
\item \Idef{org60}Inha University, Incheon, Republic of Korea
\item \Idef{org61}Institut de Physique Nucl\'{e}aire d'Orsay (IPNO), Institut National de Physique Nucl\'{e}aire et de Physique des Particules (IN2P3/CNRS), Universit\'{e} de Paris-Sud, Universit\'{e} Paris-Saclay, Orsay, France
\item \Idef{org62}Institute for Nuclear Research, Academy of Sciences, Moscow, Russia
\item \Idef{org63}Institute for Subatomic Physics, Utrecht University/Nikhef, Utrecht, Netherlands
\item \Idef{org64}Institute of Experimental Physics, Slovak Academy of Sciences, Ko\v{s}ice, Slovakia
\item \Idef{org65}Institute of Physics, Homi Bhabha National Institute, Bhubaneswar, India
\item \Idef{org66}Institute of Physics of the Czech Academy of Sciences, Prague, Czech Republic
\item \Idef{org67}Institute of Space Science (ISS), Bucharest, Romania
\item \Idef{org68}Institut f\"{u}r Kernphysik, Johann Wolfgang Goethe-Universit\"{a}t Frankfurt, Frankfurt, Germany
\item \Idef{org69}Instituto de Ciencias Nucleares, Universidad Nacional Aut\'{o}noma de M\'{e}xico, Mexico City, Mexico
\item \Idef{org70}Instituto de F\'{i}sica, Universidade Federal do Rio Grande do Sul (UFRGS), Porto Alegre, Brazil
\item \Idef{org71}Instituto de F\'{\i}sica, Universidad Nacional Aut\'{o}noma de M\'{e}xico, Mexico City, Mexico
\item \Idef{org72}iThemba LABS, National Research Foundation, Somerset West, South Africa
\item \Idef{org73}Johann-Wolfgang-Goethe Universit\"{a}t Frankfurt Institut f\"{u}r Informatik, Fachbereich Informatik und Mathematik, Frankfurt, Germany
\item \Idef{org74}Joint Institute for Nuclear Research (JINR), Dubna, Russia
\item \Idef{org75}Korea Institute of Science and Technology Information, Daejeon, Republic of Korea
\item \Idef{org76}KTO Karatay University, Konya, Turkey
\item \Idef{org77}Laboratoire de Physique Subatomique et de Cosmologie, Universit\'{e} Grenoble-Alpes, CNRS-IN2P3, Grenoble, France
\item \Idef{org78}Lawrence Berkeley National Laboratory, Berkeley, California, United States
\item \Idef{org79}Lund University Department of Physics, Division of Particle Physics, Lund, Sweden
\item \Idef{org80}Nagasaki Institute of Applied Science, Nagasaki, Japan
\item \Idef{org81}Nara Women{'}s University (NWU), Nara, Japan
\item \Idef{org82}National and Kapodistrian University of Athens, School of Science, Department of Physics , Athens, Greece
\item \Idef{org83}National Centre for Nuclear Research, Warsaw, Poland
\item \Idef{org84}National Institute of Science Education and Research, Homi Bhabha National Institute, Jatni, India
\item \Idef{org85}National Nuclear Research Center, Baku, Azerbaijan
\item \Idef{org86}National Research Centre Kurchatov Institute, Moscow, Russia
\item \Idef{org87}Niels Bohr Institute, University of Copenhagen, Copenhagen, Denmark
\item \Idef{org88}Nikhef, National institute for subatomic physics, Amsterdam, Netherlands
\item \Idef{org89}NRC Kurchatov Institute IHEP, Protvino, Russia
\item \Idef{org90}NRC «Kurchatov Institute»  - ITEP, Moscow, Russia
\item \Idef{org91}NRNU Moscow Engineering Physics Institute, Moscow, Russia
\item \Idef{org92}Nuclear Physics Group, STFC Daresbury Laboratory, Daresbury, United Kingdom
\item \Idef{org93}Nuclear Physics Institute of the Czech Academy of Sciences, \v{R}e\v{z} u Prahy, Czech Republic
\item \Idef{org94}Oak Ridge National Laboratory, Oak Ridge, Tennessee, United States
\item \Idef{org95}Ohio State University, Columbus, Ohio, United States
\item \Idef{org96}Petersburg Nuclear Physics Institute, Gatchina, Russia
\item \Idef{org97}Physics department, Faculty of science, University of Zagreb, Zagreb, Croatia
\item \Idef{org98}Physics Department, Panjab University, Chandigarh, India
\item \Idef{org99}Physics Department, University of Jammu, Jammu, India
\item \Idef{org100}Physics Department, University of Rajasthan, Jaipur, India
\item \Idef{org101}Physikalisches Institut, Eberhard-Karls-Universit\"{a}t T\"{u}bingen, T\"{u}bingen, Germany
\item \Idef{org102}Physikalisches Institut, Ruprecht-Karls-Universit\"{a}t Heidelberg, Heidelberg, Germany
\item \Idef{org103}Physik Department, Technische Universit\"{a}t M\"{u}nchen, Munich, Germany
\item \Idef{org104}Politecnico di Bari, Bari, Italy
\item \Idef{org105}Research Division and ExtreMe Matter Institute EMMI, GSI Helmholtzzentrum f\"ur Schwerionenforschung GmbH, Darmstadt, Germany
\item \Idef{org106}Rudjer Bo\v{s}kovi\'{c} Institute, Zagreb, Croatia
\item \Idef{org107}Russian Federal Nuclear Center (VNIIEF), Sarov, Russia
\item \Idef{org108}Saha Institute of Nuclear Physics, Homi Bhabha National Institute, Kolkata, India
\item \Idef{org109}School of Physics and Astronomy, University of Birmingham, Birmingham, United Kingdom
\item \Idef{org110}Secci\'{o}n F\'{\i}sica, Departamento de Ciencias, Pontificia Universidad Cat\'{o}lica del Per\'{u}, Lima, Peru
\item \Idef{org111}Shanghai Institute of Applied Physics, Shanghai, China
\item \Idef{org112}St. Petersburg State University, St. Petersburg, Russia
\item \Idef{org113}Stefan Meyer Institut f\"{u}r Subatomare Physik (SMI), Vienna, Austria
\item \Idef{org114}SUBATECH, IMT Atlantique, Universit\'{e} de Nantes, CNRS-IN2P3, Nantes, France
\item \Idef{org115}Suranaree University of Technology, Nakhon Ratchasima, Thailand
\item \Idef{org116}Technical University of Ko\v{s}ice, Ko\v{s}ice, Slovakia
\item \Idef{org117}Technische Universit\"{a}t M\"{u}nchen, Excellence Cluster 'Universe', Munich, Germany
\item \Idef{org118}The Henryk Niewodniczanski Institute of Nuclear Physics, Polish Academy of Sciences, Cracow, Poland
\item \Idef{org119}The University of Texas at Austin, Austin, Texas, United States
\item \Idef{org120}Universidad Aut\'{o}noma de Sinaloa, Culiac\'{a}n, Mexico
\item \Idef{org121}Universidade de S\~{a}o Paulo (USP), S\~{a}o Paulo, Brazil
\item \Idef{org122}Universidade Estadual de Campinas (UNICAMP), Campinas, Brazil
\item \Idef{org123}Universidade Federal do ABC, Santo Andre, Brazil
\item \Idef{org124}University of Cape Town, Cape Town, South Africa
\item \Idef{org125}University of Houston, Houston, Texas, United States
\item \Idef{org126}University of Jyv\"{a}skyl\"{a}, Jyv\"{a}skyl\"{a}, Finland
\item \Idef{org127}University of Liverpool, Liverpool, United Kingdom
\item \Idef{org128}University of Science and Techonology of China, Hefei, China
\item \Idef{org129}University of South-Eastern Norway, Tonsberg, Norway
\item \Idef{org130}University of Tennessee, Knoxville, Tennessee, United States
\item \Idef{org131}University of the Witwatersrand, Johannesburg, South Africa
\item \Idef{org132}University of Tokyo, Tokyo, Japan
\item \Idef{org133}University of Tsukuba, Tsukuba, Japan
\item \Idef{org134}Universit\'{e} Clermont Auvergne, CNRS/IN2P3, LPC, Clermont-Ferrand, France
\item \Idef{org135}Universit\'{e} de Lyon, Universit\'{e} Lyon 1, CNRS/IN2P3, IPN-Lyon, Villeurbanne, Lyon, France
\item \Idef{org136}Universit\'{e} de Strasbourg, CNRS, IPHC UMR 7178, F-67000 Strasbourg, France, Strasbourg, France
\item \Idef{org137}Universit\'{e} Paris-Saclay Centre d'Etudes de Saclay (CEA), IRFU, D\'{e}partment de Physique Nucl\'{e}aire (DPhN), Saclay, France
\item \Idef{org138}Universit\`{a} degli Studi di Foggia, Foggia, Italy
\item \Idef{org139}Universit\`{a} degli Studi di Pavia, Pavia, Italy
\item \Idef{org140}Universit\`{a} di Brescia, Brescia, Italy
\item \Idef{org141}Variable Energy Cyclotron Centre, Homi Bhabha National Institute, Kolkata, India
\item \Idef{org142}Warsaw University of Technology, Warsaw, Poland
\item \Idef{org143}Wayne State University, Detroit, Michigan, United States
\item \Idef{org144}Westf\"{a}lische Wilhelms-Universit\"{a}t M\"{u}nster, Institut f\"{u}r Kernphysik, M\"{u}nster, Germany
\item \Idef{org145}Wigner Research Centre for Physics, Hungarian Academy of Sciences, Budapest, Hungary
\item \Idef{org146}Yale University, New Haven, Connecticut, United States
\item \Idef{org147}Yonsei University, Seoul, Republic of Korea
\end{Authlist}
\endgroup

\end{document}